\documentclass[
    twoside,
    twocolumn,
    english,
    footinbib,
    tightenlines,
    nobibnotes,
    noeprint,
    aps,
    prb,                
    unsortedaddress,
    showpacs,
    superscriptaddress
]{revtex4-2}

\usepackage[range-units=single,separate-uncertainty]{siunitx}
\usepackage[color links=true,linkcolor=blue,citecolor=blue,urlcolor=blue]{hyperref}
\usepackage{relsize}

\usepackage{graphicx}
\usepackage{svg}
\usepackage{xcolor}
\usepackage{orcidlink}
\usepackage{enumitem}
\usepackage{ulem}

\usepackage{multirow}

\usepackage{amsmath,amsfonts,amssymb,mathrsfs,amsthm}
\usepackage{mathtools,braket}
\usepackage{bm,bbm}

\usepackage{comment}
\usepackage{tikz}

\newcolumntype{P}[1]{>{\centering\arraybackslash}p{#1}}

\hypersetup{
    colorlinks,
    linkcolor={blue},
    citecolor={blue},
    urlcolor={blue}
}

\newcommand{\ii}{\mathrm{i}}
\renewcommand{\i}{\mathrm{i}}

\renewcommand{\Im}{\mathrm{Im}}

\def\ie{{i.e.},\ }

\def\cf{{cf.}\ }

\def\appendixname{Appendix}

\def\XXint#1#2#3{{\setbox0=\hbox{$#1{#2#3}{\int}$}
     \vcenter{\hbox{$#2#3$}}\kern-.5\wd0}}

\begin{document}


\title{Chiral Gapless Spin Liquid in Hyperbolic Space}

\author{Felix Dusel\orcidlink{0000-0002-5082-8370}}
\affiliation{Institut für Theoretische Physik und Astrophysik and Würzburg-Dresden Cluster of Excellence ct.qmat, Julius-Maximilians-Universität, 97074 Würzburg, Germany}

\author{Tobias Hofmann\orcidlink{0000-0002-1888-9464}}
\email{Corresponding author: tobias.hofmann@uni-wuerzburg.de}
\affiliation{Institut für Theoretische Physik und Astrophysik and Würzburg-Dresden Cluster of Excellence ct.qmat, Julius-Maximilians-Universität, 97074 Würzburg, Germany}

\author{Atanu Maity\orcidlink{0000-0001-7822-124X}}
\affiliation{Institut für Theoretische Physik und Astrophysik and Würzburg-Dresden Cluster of Excellence ct.qmat, Julius-Maximilians-Universität, 97074 Würzburg, Germany}

\author{R\'emy Mosseri\orcidlink{0000-0002-2053-7724}}
\affiliation{Sorbonne Universit\'e, CNRS, Laboratoire de Physique Th\'eorique de la Mati\`ere Condens\'ee, LPTMC, F-75005 Paris, France}

\author{Julien Vidal\orcidlink{0000-0002-1788-643X}}
\affiliation{Sorbonne Universit\'e, CNRS, Laboratoire de Physique Th\'eorique de la Mati\`ere Condens\'ee, LPTMC, F-75005 Paris, France}
\affiliation{Department of Physics and Quantum Center for Diamond and Emergent Materials (QuCenDiEM), Indian Institute of Technology Madras, Chennai 600036, India}

\author{Yasir Iqbal\orcidlink{0000-0002-3387-0120}}
\affiliation{Department of Physics and Quantum Center for Diamond and Emergent Materials (QuCenDiEM), Indian Institute of Technology Madras, Chennai 600036, India}

\author{Martin Greiter\orcidlink{0000-0003-2008-4013}}
\affiliation{Institut für Theoretische Physik und Astrophysik and Würzburg-Dresden Cluster of Excellence ct.qmat, Julius-Maximilians-Universität, 97074 Würzburg, Germany}

\author{Ronny Thomale\orcidlink{0000-0002-3979-8836}}
\affiliation{Institut für Theoretische Physik und Astrophysik and Würzburg-Dresden Cluster of Excellence ct.qmat, Julius-Maximilians-Universität, 97074 Würzburg, Germany}
\affiliation{Department of Physics and Quantum Center for Diamond and Emergent Materials (QuCenDiEM), Indian Institute of Technology Madras, Chennai 600036, India}

\date{\today}


\begin{abstract}
We analyze the Kitaev model on the $\{9,3\}$ hyperbolic lattice. The $\{9,3\}$ is  formed by a regular tricoordinated tiling of nonagons, where the three-color coding of bonds according to the inequivalent Kitaev Ising spin couplings yields the natural generalization of the original Kitaev model for Euclidean regular honeycomb tiling. Upon investigation of the bulk spectrum for large finite size droplets, we identify a gapless chiral $\mathbb{Z}_2$ spin liquid state featuring spontaneous time-reversal symmetry breaking. Because of to its noncommutative translation group structure, such type of hyperbolic spin liquid is conjectured to feature chiral quasiparticles with a potentially non-Abelian Bloch profile. 
\end{abstract}


\maketitle


\begin{figure*}
  \centering
  \begin{tikzpicture}
      \node[anchor=south west] at (12, 0) {
          \includegraphics[
              height=4.2cm,
              trim={0.2cm 0.2cm 0.2cm 0.2cm},
              clip
              ]{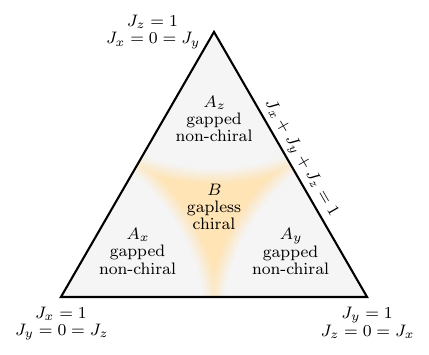}
      };
      \node[anchor=south west] at (0, 0) {
          \includegraphics[
              height=4.2cm,
              trim={0.4cm 0.4cm 0.2cm 0.4cm},
              clip
              ]{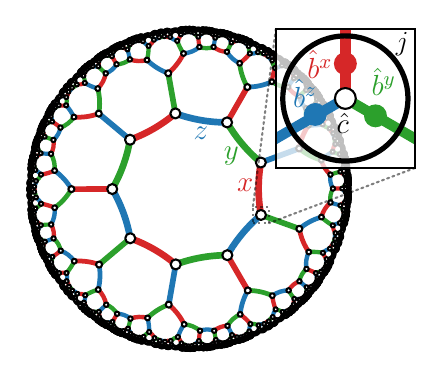}
      };
      \node[anchor=west] at (0, 4.2) {\textbf{(a)}};
      \node[anchor=south west] at (6.5, 0) {
          \includegraphics[
              height=4.2cm,
              trim={0.5cm 0.5cm 0.5cm 0.5cm},
              clip
              ]{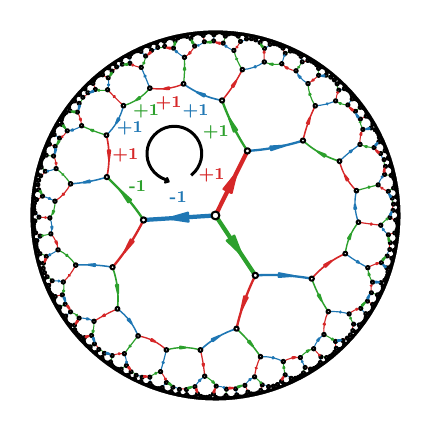}
      };
      \node[anchor=west] at (6.5, 4.2) {\textbf{(b)}};
      \node[anchor=west] at (12, 4.2) {\textbf{(c)}};
  \end{tikzpicture}
  \caption{
    \label{fig:model}
    \textbf{(a)} 
    Droplet of the $\{9, 3\}$ hyperbolic lattice with Kitaev-colored bonds for each coupling $J_x, J_y, J_z$. The inset shows schematically the four Majorana fermions $\hat{b}_j^\alpha, \hat{c}_j$ introduced at site $j$.
    \textbf{(b)} 
    Direction convention for each link $\hat{u}_{j k}$. If all eigenvalues $u_{j k}$ are $\pm1$, the $\pm\pi/2$ flux sector is realized.
    \textbf{(c)}
    Schematic phase diagram in the $J_x + J_y + J_z = 1$ plane. 
    }
\end{figure*}


\paragraph*{Introduction.}
\label{sec:intro}

Quantum spin liquids have been in the focus
of contemporary research on strongly correlated electron systems for
five decades~\cite{Savary_2017}. In experiment, the unambiguous identification of such
an intricate quantum fluid state of matter is notoriously difficult, and rarely, if ever,
possible beyond reasonable doubt~\cite{doi:10.1126/science.aay0668}. In theory, our understanding has
progressed at various frontiers, yet at highly different rates. Due to
the seminal work by Wen~\cite{PhysRevB.65.165113}, an assumingly exhaustive classification of
quantum spin liquids is given from their associated emergent gauge
structure at low energies. There usually is, however, still a long way
to go from quantum order classifications, and their related parton
mean fields, to a refined
microscopic realization and stabilization of such states. This is why,
in particular for quantum spin liquids and quantum many-body phases in
general, it is highly desirable to accomplish parent Hamiltonians
whose ground state unambiguously matches such a state. Typically, a combination of optimization
ansatz and educated guess is employed to scan a multidimensional
Hamiltonian parameter space~\cite{Qi2019determininglocal,PhysRevX.8.031029,PhysRevB.98.081113}. 

The Kitaev paradigm of spin liquid parent Hamiltonians has proven
particularly successful~\cite{KITAEV20062}. Conceived as a
bond-dependent anisotropic Ising Hamiltonian between neighboring sites on
the honeycomb lattice, it has brought about various exactly solvable
instances of quantum spin liquids with an emergent $\mathbb{Z}_2$
gauge structure. In terms of spin liquids featuring
topological order, it includes the gapped Abelian toric code liquid and,
upon consideration of an external magnetic field~\cite{KITAEV20062} or through
spontaneous time-reversal (T) symmetry breaking via lattice graphs with odd loops~\cite{PhysRevLett.99.247203,PhysRevB.101.041114}, a gapped non-Abelian
chiral Majorana liquid with spinon quasiparticles which resemble
Pfaffian liquid quasiparticles~\cite{MOORE1991362,PhysRevLett.102.207203}. 
The Kitaev model further harbors the gapless $\mathbb{Z}_2$ spin liquid centered around isotropic Ising couplings. Further progress has been reached by
Hermanns and Trebst~\cite{PhysRevB.89.235102} who first employed that if a
unique bond attribution to three subsets can be made for a
semiregular tricoordinated tiling, this still yields exact
solvability of the Kitaev model. Upon transcending from two to three
spatial dimensions, this enabled them to define Kitaev spin liquid states
with a (potentially unstable~\cite{PhysRevLett.115.177205}) spinon Fermi surface as well as chiral spin liquid states with
crystalline gauge ordering of $\mathbb{Z}_2$ fluxes~\cite{PhysRevB.101.045118}. 

Since the nature of the lattice strongly affects the magnetic frustration profile, any parent
Hamiltonian search for quantum spin liquids seeks to consider all
variations of potentially interesting lattices. Conventionally, this
implies to be constrained to tessellations of Euclidean flat
space. Recent progress on hyperbolic matter, however, has broadened
the view to also consider Hamiltonians operating on negatively curved
spaces and lattice tessellations
thereof~\cite{PhysRevB.105.125118}. While the field initially started
from unconventional noninteracting states and their description in
terms of symmetry protected topological phases~\cite{PhysRevLett.129.246402,PhysRevLett.132.206601,PhysRevLett.128.166402} and hyperbolic band
theory~\cite{doi:10.1126/sciadv.abe9170,PhysRevLett.129.088002,maciejkorayan,PhysRevLett.131.226401}, it is now starting to reach the realms of quantum many-body
physics through metal-insulator transitions, charge density wave states, and magnetism~\cite{goetz2024hubbardheisenbergmodelshyperbolic}, fractional Chern insulators~\cite{he2024hyperbolicfractionalcherninsulators},  and lattice-regularized
formulations of AdS-CFT correspondence~\cite{basteiro2023breitenlohner,dey2024simulatingholographicconformalfield}. An immediate platform for tabletop experimental realizations
of hyperbolic space presents itself to be quantum and classical circuit networks~\cite{kollar,Lenggenhager-nc-2022,Chen-nc-2023}.

In this Letter, we introduce the notion of two-dimensional (2D) hyperbolic spin
liquids in the framework of a generalized Kitaev model. Applying the same logical
thread as in Euclidean space, the focus of consideration lies on
tricoordinated lattice tessellations in the hope of reaching an exactly
solvable Hamiltonian. While in principle any of the vast hyperbolic
space tessellations that fulfill this tricoordination deserves further
investigation, we will particularize to a microscopic model for an
exotic spin liquid state that features gapless excitations, and yet
spontaneously breaks time reversal symmetry. Notwithstanding previous conjectures about a possible chiral gapless spin liquid on the kagome lattice~\cite{PhysRevB.92.060407,SciPostPhys.4.1.004}, the microscopic setting and the exact ground state we find has no immediate analogue in Euclidean space, and hence expands the hitherto known landscape of spin liquids to uncharted territory.

Realizing gapless excitations for a Kitaev Hamiltonian is intimately
tied to regular tiling, \ie tessellations with only one type of
regular polygon, while spontaneous T~symmetry breaking requires
odd-sided loops. Such a tessellation of 2D Euclidean space by regular
odd-sided tilings does not exist. Instead, one would have to resort to
semi-regular tilings such as the Fisher star
lattice~\cite{PhysRevLett.99.247203,PhysRevB.76.180404,PhysRevB.78.125102} as well as
non-Archimedean~\cite{PhysRevB.101.041114,PhysRevB.76.180404} and amorphous~\cite{Cassella2023} realizations, all of which feature gapped chiral Kitaev spin liquids. This motivates us to transcend to 2D hyperbolic space to realize tricoordinated, regular odd-loop tilings. Among the vast tessellations that fulfill the
constraint, the $\{9,3\}$
lattice Fig.~\ref{fig:model}(a)\ stands out, as it features the smallest loop size of hyperbolic tiling whose bond coloring appears highly canonical, and even suggests itself to be the most natural generalization of the Euclidean honeycomb Kitaev model. Our analysis of the $\{9,3\}$ hyperbolic Kitaev model finds a gapless state which, due to the presence of odd loops, features
spontaneous time-reversal symmetry breaking. We thus show the Kitaev
model to harbor a gapless, chiral $\mathbb{Z}_2$ spin liquid state
whose nature of excitations, however, is conjectured to be intimately
dependent on the non-commutative translation group inherent to
hyperbolic space.

\begin{figure*}
  \centering
  \begin{tikzpicture}
      \node[anchor=south west] at (0, 0) {
          \includegraphics[
              scale=0.87,
              trim={0 0.15cm 0 0},
              clip
              ]{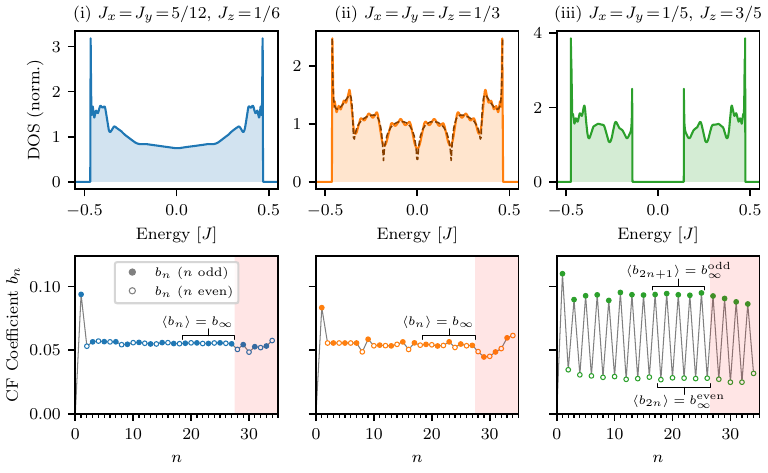}
      };
      \node[anchor=south west] at (11.9, 2.7) {
          \includegraphics[
              scale=0.78,
              trim={0 0.2cm 0 0.35cm},
              clip
              ]{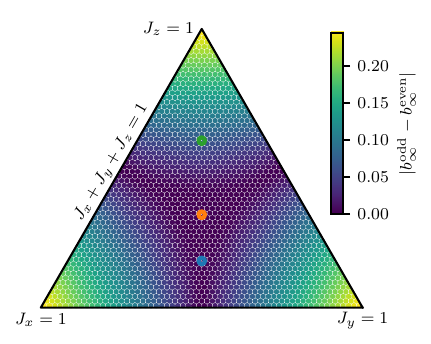}
      };
      \node[anchor=south west] at (11.9, 0) {
          \includegraphics[
              scale=0.78,
              trim={0 0.3cm 0 0},
              clip
              ]{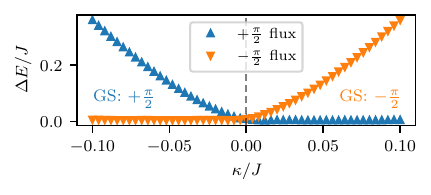}
      };
      
      \node[anchor=west] at (0, 6.7) {\textbf{(a)}};
      \node[anchor=west] at (11.9, 6.7) {\textbf{(b)}};
      \node[anchor=north west] at (11.9, 2.6) {\textbf{(c)}};
  \end{tikzpicture}

  \caption{
    \label{fig:phase_diagram_dos}
    \textbf{(a)}~ 
    Density of states (DOS) of~\eqref{eq:SingleParticleHamiltonian} in the $\pi/2$ flux sector computed by the continued fraction (CF) method on a droplet with \num{4.2e6} sites for representative points in the phase diagram. 
    In (ii) we compare the result to the DOS computed by exact diagonalization of a PBC cluster (dashed line).
    The lower panels show the corresponding CF coefficients $b_n$ expected to converge to a single value $b_{\infty}$ in a gapless phase, yet exhibiting two attractors $b_{\infty}^{\mathrm{odd}}$ and $b_{\infty}^{\mathrm{even}}$ for a spectral gap.
    Because of finite-size effects, the coefficients for $n > 28$ are discarded (red-shaded region). As the CF expansion is not yet converged, small ripple artifacts are observed on top of the DOS. For the termination term we use an average over nine coefficients.
    \textbf{(b)}~
    Phase diagram for $|b_{\infty}^{\mathrm{odd}} - b_{\infty}^{\mathrm{even}}|$. It is vanishing (finite) within phase $B$ ($A$). The points of (a)\ are marked by their colors.
    \textbf{(c)}~
    Fermion gap versus next-nearest neighbor coupling $\kappa$ for $J_x = J_y = J_z = \frac{1}{3}$ and an \num{8436} site PBC cluster. The degeneracy of the $\pm \frac{\pi}{2}$ flux sectors is lifted. The respective ground state sector exhibits a gap.
    }
\end{figure*}


\paragraph*{Model.}

We consider the Kitaev Hamiltonian 
\begin{align}\label{eq:KitaevHamiltonian}
  \hat{H} 
  = 
  -J_{x} \! \sum_{\langle j, k\rangle_x} \hat{\sigma}_j^x\hat{\sigma}_k^x 
  -J_{y} \! \sum_{\langle j, k\rangle_y} \hat{\sigma}_j^y\hat{\sigma}_k^y 
  -J_{z} \! \sum_{\langle j, k\rangle_z} \hat{\sigma}_j^z\hat{\sigma}_k^z
\end{align}
on a three-colored $\{9, 3\}$ lattice. There, $\hat{\sigma}_{i}^{\alpha}$ denote Pauli matrices acting on site $i$, and $J_{\alpha} > 0$ are the coupling constants along the respective link of type $\alpha = x, y, z$. 
The different link types are shown in Fig.~\ref{fig:model}(a)\ as red, green and blue lines, respectively.
Although model~\eqref{eq:KitaevHamiltonian} can be solved exactly for all tricoordinated three-colorable hyperbolic lattices, the $\{9, 3\}$ appears to be the natural hyperbolic extension of the original Euclidean model due to the regular coloring scheme of repeated $x$, $y$, and $z$ links upon going counterclockwise around each plaquette (Fig.~\ref{fig:model}). Invoking $(p-2)(q-2)>4$ obeyed by all hyperbolic tilings, this canonical coloring scheme exists for all $\{p, 3\}$ lattices with $p \equiv 0 \mod 3, \; p>6 $, so that the $\{9, 3\}$ lattice is the one with smallest loop size.

The Hamiltonian~\eqref{eq:KitaevHamiltonian} possesses a set of conserved quantities, one associated with each nonagon plaquette, given by the mutually commuting Wilson loop operators $\hat{W}_{p} = \hat{K}_{12} \hat{K}_{23} \hat{K}_{34} \hat{K}_{45} \hat{K}_{56} \hat{K}_{67} \hat{K}_{78} \hat{K}_{89} \hat{K}_{91}$, where $\hat{K}_{i j} = \hat{\sigma}_{i}^{\alpha} \hat{\sigma}_{j}^{\alpha}$, and the sites $i$ and $j$ are labeled by indices $i,j =1,\cdots,9$ counterclockwise around a plaquette and connected by a bond of type $\alpha$. Since all plaquette operators commute, the Hamiltonian attains a block diagonal form according to the $\hat{W}_p$ eigenvalues. 
In contrast to Kitaev models with even-loop plaquettes, these are given by $\pm \ii$ rather than $\pm 1$, implying 
fictitious magnetic flux of $\pm \pi/2 $. Equation~\eqref{eq:KitaevHamiltonian} is invariant under $\mathrm{T}$, which is an antiunitary symmetry for spin $1/2$. The eigenvalues of $\hat{W}_p$ change sign under $\mathrm{T}$, thus forming Kramers pairs and rendering the overall spectrum twofold degenerate.


\paragraph*{Majorana fermion representation.}

Following Kitaev, we transform~\eqref{eq:KitaevHamiltonian} into its Majorana representation.
At each site, we express the spin operator by four Majorana operators $\hat{c}_j, \hat{b}_j^\alpha$, according to $\hat{\sigma}_{j}^{\alpha} = \ii \hat{b}_j^\alpha \hat{c}_j$, which yields
\begin{align}
\label{eq:SingleParticleHamiltonian}
  \hat{H} = \frac{\ii}{2} \sum_{j, k} J_{\alpha_{jk}} \hat{u}_{jk} \hat{c}_j \hat{c}_k ,
\end{align}
where we introduced the Majorana bilinears $\hat{u}_{j k} = \i \hat{b}_j^\alpha \hat{b}_k^\alpha$. These link operators are antisymmetric under link inversion $\hat{u}_{j k} = -\hat{u}_{k j}$ and under $\mathrm{T}$, $\hat{u}_{j k} \overset{\mathrm{T}}{\longrightarrow} 
\hat{u}_{k j}$.
The Majorana transformation effectively doubles the Hilbert space. Defining $\hat{D}_i = \ii \hat{b}_i^x \hat{b}_i^y \hat{b}_i^z \hat{c}_i$ and $\hat{D}_i \ket{\psi} = +1\ket{\psi}\;\forall i$, the states $\{\ket{\psi}\}$ form the physical Majorana subspace.

Since each link couples only one flavor of Majoranas, the operators $\hat{u}_{j k}$ commute with each other and Eq.~\eqref{eq:SingleParticleHamiltonian}. Therefore, we can replace them by the corresponding eigenvalues $u_{j k} = \pm 1$ relating to the $\pm \pi/2$ sectors interpreted as a $\mathbb{Z}_2$ gauge field. Once a configuration of these eigenvalues is specified, the Hamiltonian~\eqref{eq:SingleParticleHamiltonian} takes a noninteracting single-particle form. 
In order to express the gauge invariant eigenvalues $W_p$ of the Wilson loop operators by $u_{j k}$, we fix a direction for all links according to Fig.~\ref{fig:model}(b).
Then, the Majorana transformation implies $W_p = (-\ii)^9 \prod_{\langle j k \rangle_{p}} u_{j k}$, where the product runs in a counterclockwise cycle over all links of a given plaquette~$p$.


\paragraph*{Ground state.}

Given a flux configuration, the energy of an eigenstate of \eqref{eq:KitaevHamiltonian} is computed by the sum over the absolute value of all eigenvalues of~\eqref{eq:SingleParticleHamiltonian}. For bipartite (and hence even-loop) $\{p, q\}$ lattices Lieb's theorem~\cite{Lieb1994} is applicable, stating that the ground state lies in the uniform flux-$0$ or flux-$\pi$ sector. For the $\{9, 3\}$ lattice, a recent conjecture~\cite{Cassella2023} as well as our numerical results (see Supplemental Material~\cite{appendix}, Sec.~A) indicate that the ground state(s) are characterized by uniform vortex decorations as well, such that the two-dimensional ground state manifold is constituted by the uniform $\pm \pi/2$ flux sectors, which form a Kramers doublet.


We calculate the bulk density of states (DOS) of the Majorana Hamiltonian using the continued fraction (CF) method~\cite{mosseri2023density} (see~\cite{appendix}, Sec.~B). Representative points of parameter space are depicted in Fig.~\ref{fig:phase_diagram_dos}(a). 
Although the CF expansion is not yet fully converged for the available system sizes, we identify two distinct regimes by the number of attractors of the CF coefficients:
a gapped phase $A$ and a gapless domain $B$ (Fig.~\ref{fig:phase_diagram_dos}(b)). Furthermore, we find the phase boundary approximately along the lines where one coupling constant is equal to the sum of the other two. Note that the symmetry of the phase diagram is linked to the canonical colorability of the $\{9,3\}$ lattice.


\paragraph*{Dimerized phase A.}

\begin{figure}
  \centering
  \begin{tikzpicture}
    \node[anchor=south west] at (0, 0) {
      \includegraphics[width=0.98\columnwidth]{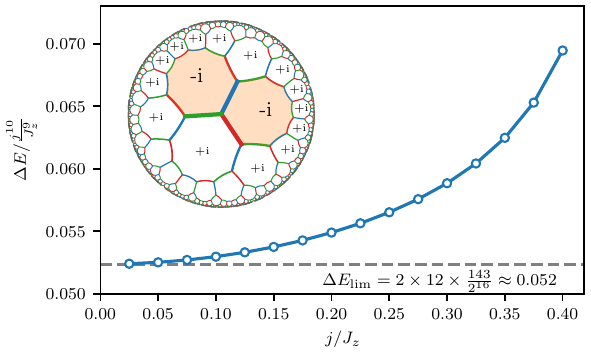}
              };
  \end{tikzpicture}
  \caption{
    \label{fig:perturbation_theory}
    Energy increase $\Delta E$ due to vison pair creation separated by a $z$-type bond for $J_x = J_y = j \ll J_z$ in a \num{10024} site OBC cluster. The analytical value $\Delta E_{\textrm{lim}}$ predicted by Eq.~\eqref{eq:H_eff} is recovered for $j\rightarrow 0$. 
  }
\end{figure}

The three regions $A_x$, $A_y$ and $A_z$, \ie the dimerized phase regimes dominated by either strong Ising coupling, are equivalent modulo Ising axis rotation. Although the CF method only converges slowly approaching the phase boundary to the gapless $B$ domain, the band gap can be clearly identified. 
In order to further analyze the $A$ phases we consider the dimerized limit of strong coupling anisotropy $(J_x,J_y,J_z)=(0,0,J)$ where the system reduces to decoupled Ising dimers.
We contract the Ising dimers to a point, and regard them as effective spin-1/2 degrees of freedom on the $\{6,4\}$ hyperbolic lattice. There are fermionic and $\mathbb{Z}_2$ vortex excitations: A fermion is created when we break the alignment of two spins on a strong bond, which costs energy $\sim\!2 J_z$. The low-energy physics of the $\mathbb{Z}_2$ vortices is governed by the effective two-plaquette~\cite{PhysRevLett.100.057208,PhysRevB.78.245121} Hamiltonian
\begin{align}
  \label{eq:H_eff}
  \hat{H}_\text{eff}^{(10)} =&\,
  \text{const} 
  + 
  \frac{143 \, J_x^4 \, J_y^4}{2^{16} J_z^9} \nonumber\\
  &\times \bigg( 
    J_y^2 \sum_{\langle p, p' \rangle_{x}} \hat{W}_p \hat{W}_{p'} 
  + J_x^2 \sum_{\langle p, p' \rangle_{y}} \hat{W}_p \hat{W}_{p'}   
  \bigg) ,
\end{align}
where $\hat{W}_p$ and $\hat{W}_{p'}$ are Wilson loop operators on two neighboring nonagons, sharing a weak $x$ or $y$ bond (see~\cite{appendix}, Sec.~C).
Flipping a bond $u_{j k} \to - u_{j k}$ creates a pair of visons in the two adjacent nonagons, which increases the energy of the state by $\Delta E$ as the new $-\frac{\pi}{2}$ plaquettes are surrounded by $+\frac{\pi}{2}$ fluxes. Figure~\ref{fig:perturbation_theory} compares $\Delta E$ observed by exact diagonalization of an OBC cluster with the result of Eq.~\eqref{eq:H_eff}. This effective model is a hyperbolic version of the Kitaev Toric Code~\cite{breuckmann2017hyperbolic}. We do not further elaborate on the region $A$, as it largely corresponds to its Euclidean analog. A subtle issue arises for topologically ordered ground states on hyperbolic lattices regarding the topological ground state degeneracy. The compactified hyperbolic space suggests a higher Riemann manifold genus which grows with system size. This implies a challenge in the reconciliation with the assumed relation between topological ground state degeneracy and genus~\cite{PhysRevB.41.9377}, a principal issue which will be addressed elsewhere~\cite{inprep}.

\begin{figure*}
  \begin{tikzpicture}
      \node[anchor=north west] at (0, -0.25) {
          \includegraphics[trim={0.2cm 0.2cm 1cm 0.2cm},clip,width=6cm]{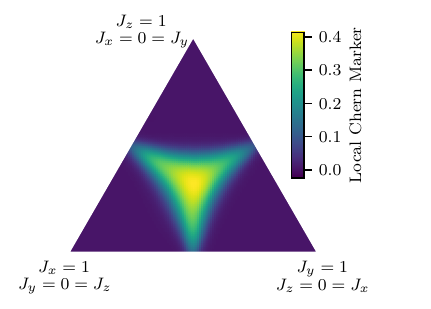}
      };
      
      \begin{scope}[shift={(6.2cm+0*2.8cm, 0.1cm)}]
          \clip (3.3/2+0.1, -3.3/2-0.1) circle (1.7);
          \node[anchor=north west] at (0, 0) {
              \includegraphics[trim={0.5cm 0.5cm 0.5cm 0.5cm},clip,width=3.3cm]{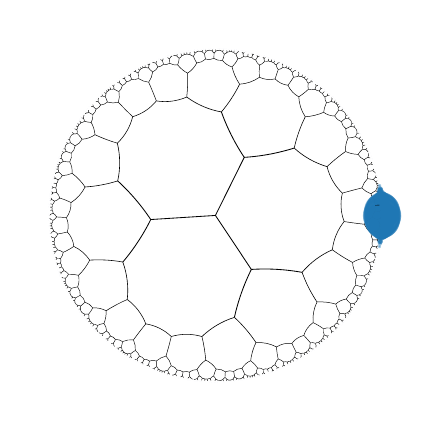}
          };
      \end{scope}
      \node[anchor=north] at (6.2+0*2.85+3.3/2+0.1, -3.2) {$t = 0$};  
      
      \begin{scope}[shift={(6.2cm+1*2.8cm,-1.9cm)}]
          \clip (3.3/2+0.1, -3.3/2-0.1) circle (1.7);
          \node[anchor=north west] at (0, 0) {
              \includegraphics[trim={0.5cm 0.5cm 0.5cm 0.5cm},clip,width=3.3cm]{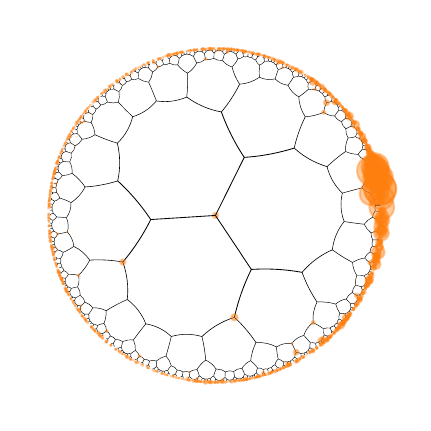}
          };
      \end{scope}
      \node[anchor=south] at (6.2+1*2.85+3.3/2+0.1,-2.1) {$t = 800$};  

      \begin{scope}[shift={(6.2cm+2*2.8cm, 0.1cm)}]
          \clip (3.3/2+0.1, -3.3/2-0.1) circle (1.7);
          \node[anchor=north west] at (0, 0) {
              \includegraphics[trim={0.5cm 0.5cm 0.5cm 0.5cm},clip,width=3.3cm]{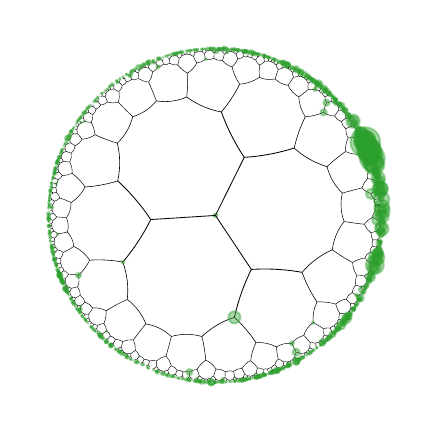}
          };
      \end{scope}
      \node[anchor=north] at (6.2+2*2.8+3.3/2+0.1, -3.2) {$t = 1600$};

      \begin{scope}[shift={(6cm+3*2.8cm, -1.9cm)}]
          \clip (3.3/2+0.1, -3.3/2-0.1) circle (1.7);
          \node[anchor=north west] at (0, 0) {
              \includegraphics[trim={0.5cm 0.5cm 0.5cm 0.5cm},clip,width=3.3cm]{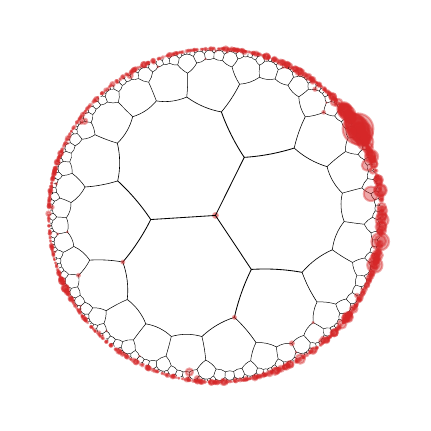}
          };
      \end{scope}
      \node[anchor=south] at (6.2+3*2.8+3.3/2+0.1, -2.1) {$t = 2400$};  
      
      \node[anchor=north west] at (0, 0) {\textbf{(a)}};
      \node[anchor=north west] at (6.2, 0) {\textbf{(b)}};
      \node[rectangle,draw,anchor=north east, align=right] at (\linewidth-0.2cm, -0.1cm) {\footnotesize System size:\\\num{10024} sites};
  \end{tikzpicture}
  \caption{
      \label{fig:phase_diagram_chern}
      \textbf{(a)}
      Phase diagram of the local Chern marker (LCM, cf.~Eq.~\eqref{eq:chern_marker}) in the plane $J_x + J_y + J_z = 1$ for a droplet with 1575 sites. The LCM is calculated from~\eqref{eq:SingleParticleHamiltonian} including all states with negative energy. It is finite but not quantized in the chiral phase $B$, and vanishes in the $A$ phases.
      \textbf{(b)}
      Time evolution under~\eqref{eq:SingleParticleHamiltonian} of a wave packet initially localized at the edge in the uniform $\pi/2$ flux sector for $J_x = J_y = J_z = \frac{1}{3}$. The center of the packet propagates unidirectionally along the boundary. The area of each circle is proportional to the probability to measure the excitation at the corresponding site.
  }
\end{figure*}

\paragraph*{Chiral gapless phase B.}

Panels (i), (ii) of Fig.~\ref{fig:phase_diagram_dos}(a)\ show the DOS in the gapless regime for representative points in parameter space. 
In contrast to the gapless regime for the Euclidean Kitaev model, our bulk spectra analysis indicates a finite, nonvanishing DOS at zero energy. This suggests a connection to the generically deviating character of hyperbolic semimetals~\cite{PhysRevLett.132.206601}, where the noncommutative geometry implies an effective Bloch band theory with higher-dimensional irreducible representations of translation. This deserves further investigation for the $\{9,3\}$ lattice in general.
Moreover, the gapless phase we find here is intrinsically chiral since, due to the odd loops, it spontaneously breaks T symmetry in the thermodynamic limit. 
For flakes up to $10^5$ sites, the Majorana representation~\eqref{eq:SingleParticleHamiltonian} can be diagonalized numerically~\cite{schrauth2023hypertiling}.
To investigate the chiral character of phase $B$, we calculate the local Chern marker (LCM) proposed by Kitaev~\cite{KITAEV20062,PhysRevB.106.155120},
\begin{align}\label{eq:chern_marker}
  C = 12 \pi \ii \sum_{j \in A} \sum_{k \in B} \sum_{l \in C} (\hat{P}_{jk}\hat{P}_{kl}\hat{P}_{lj}-\hat{P}_{jl}\hat{P}_{lk}\hat{P}_{kj}),
\end{align}
where $\hat{P}$ is the projector onto states with negative energy; $A$, $B$, and $C$ denote three sectors in counterclockwise order that evenly partition the hyperbolic flake not including boundary sites (see~\cite{appendix}, Sec.~E). We find the LCM to vanish everywhere in the dimer phase $A$, but to be finite valued in the gapless phase for either one of the $\pm \pi/2$ flux sectors (\cf Fig.~\ref{fig:phase_diagram_chern}(a)). We observe that the value of the LCM is not quantized. This is not inconsistent with the expected phenomenology, since the Chern number of a topological band is only quantized if the Fermi level resides within a band gap. Furthermore, the LCM calculation might be affected due to finite-size or edge effects in the hyperbolic plane.

In order to investigate the chiral character of the gapless phase $B$ from a different angle, we conduct a numerical propagation experiment in the Majorana representation of~\eqref{eq:KitaevHamiltonian}. We apply the time evolution of~\eqref{eq:SingleParticleHamiltonian} in the $+\pi/2$ flux configuration to an initial state with energy expectation value $\langle H \rangle = 0$, which represents an excitation localized at the edge of a droplet (Fig.~\ref{fig:phase_diagram_chern}(b), see~\cite{appendix}, Sec.~F). After $800$, $1600$, and $2400$ time steps we observe that the excitation has moved unidirectionally along the edge. As expected, the direction of propagation is reversed for the $-\pi/2$ flux sector.


\paragraph*{Perturbative magnetic field.}
Under the application of a small magnetic field, the Hamiltonian becomes $\hat{\widetilde{H}} = \hat{H} + \hat{H'}$ with
\begin{align}    
  \hat{H}' =
  -\kappa \sum_{\substack{(j,k,l)_{\alpha\gamma}\\\beta\neq\alpha, \beta\neq\gamma}} \hat{\sigma}_j^{\alpha} \, \hat{\sigma}_k^{\beta} \, \hat{\sigma}_l^{\gamma} .
\end{align}
The sum runs over all three-tuples of sites $(j,k,l)_{\alpha\gamma}$, such that sites $j$ and $k$ are connected anticlockwise around a plaquette by an $\alpha$-type bond, and $k$ and $l$ by a $\gamma$-type bond. $\beta$ is then given by the third bond flavor. 
This simple form is sustained by the regular coloring characteristic for the $\{9, 3\}$ lattice.
In the Majorana language the Hamiltonian takes the form 
\begin{align}
  \hat{H}' = \mathrm{i} \, \frac{\kappa}{2} \sum_{(j,k,l)} \hat{u}_{j k} \hat{u}_{k l} \hat{c}_j \hat{c}_l ,
\end{align}
which corresponds to a second-nearest neighbor hopping term for the Majorana fermions.

We set $J_x = J_y = J_z = \frac{1}{3}$ and analyze the model for $\kappa \neq 0$. As the magnetic term breaks time-reversal symmetry, the $\pm \pi/2$ flux sectors are no longer degenerate. For $\kappa > 0$, the ground state is given by the uniform $-\pi/2$ configuration, while the $+\pi/2$ configuration constitutes the ground state for $\kappa < 0$. 
We compute the fermion gap for both sectors (Fig.~\ref{fig:phase_diagram_dos}~(c)), and find that a gap opens for arbitrarily small $\kappa$.


\paragraph*{Conclusion and outlook.} 

As a future endeavor, it will be important to study the behavior of the spin-spin correlation function of the gapless chiral spin liquid and obtain the dynamic exponent. This would provide insights into the precise nature of this ground state. 
From here, utilizing a projective symmetry group (PSG) classification of spin liquids towards characterizing the microscopic symmetries of the underlying quantum orders in hyperbolic space presents itself as a natural thread of investigation. This is because the PSG may enable us to identify the symmetries protecting the gapless chiral excitations, and to study their intertwining with a potentially non-Abelian Bloch translation group profile.


\paragraph*{Note added.} 

Recently, we
became aware of an independent simultaneous work addressing hyperbolic Kitaev spin liquids on the $\{8,3\}$ lattice with and without the presence of an external magnetic field~\cite{lenggenhager2024hyperbolicspinliquids}.


\paragraph*{Acknowledgements.}

We thank A. Stegmaier, I. Boettcher, P. M. Lenggenhager, S. Dey, T. Bzdušek, and J. Maciejko for discussions, as well as R. Vogeler for providing a large PBC cluster of the $\{9, 3\}$ hyperbolic lattice.
The work is funded by the
Deutsche Forschungsgemeinschaft (DFG, German Research Foundation)
through Project-ID 258499086 - SFB 1170 and through the
W\"urzburg-Dresden Cluster of Excellence on Complexity and Topology in
Quantum Matter -- \textit{ct.qmat} Project-ID 390858490 - EXC 2147 as well as 
by the National Science Foundation through NSF-PHY-2210452.


\paragraph*{Author contributions.}
R.T. initiated and supervised the project. F.D., T.H., and R.T. designed the model system. F.D. and T.H. carried out the calculations with discursive and calculation input from J.V., R.M., A.M., M.G., and Y.I.. All authors contributed to writing the manuscript.


%


\def\appendixname{Supplemental Material}

\clearpage
\numberwithin{equation}{section}
\numberwithin{figure}{section}
\appendix
\setcounter{page}{1}
\onecolumngrid

{
    \centering 
    \textbf{SUPPLEMENTAL MATERIAL}

}


\section{Ground state flux sector and excitations}
\label{appx:flux_sector_investigation}

The Kitaev Hamiltonian (Eq.~\eqref{eq:KitaevHamiltonian} of the main text) commutes with the plaquette operators $\hat{W}_p$.
Therefore, the Hilbert space can be decomposed into the eigenspaces of $\hat{W}_{p}$, which we call flux sectors. They are labeled by the eigenvalues of each plaquette operator $W_p = \pm \i$. 
The eigenvalues $W_p$ can be expressed in terms of the eigenvalues of the link operators, $u_{j k} = \pm 1$.
A given flux configuration on the plaquettes of a lattice can arise from multiple distinct link configurations. If all $u_{j k} = +1$, one finds $W_p = -\i \;\forall p$, \ie each plaquette carries a flux of $-\frac{\pi}{2}$. 
Accordingly, we call this sector the uniform $-\pi/2$ flux sector. If all $u_{j k} = -1$, the plaquette eigenvalues are $W_p = +\i \;\forall p$. Then each plaquette carries a flux of $\frac{\pi}{2}$. Both map onto each other under time reversal, and form a Kramers pair.

We wish to identify the flux sector in which the ground state is located. 
For bipartite lattices Lieb's theorem~\cite{Lieb1994} can be employed. 
Regarding lattices consisting of regular $p$-gons it states that if the graph is reflection symmetric with respect to an axis that does not intersect with any lattice site, the ground state always lies in either the sector where all plaquettes have flux $0$ or flux $\pi$. If $p=2 \mod 4$ (as for the Kitaev Honeycomb Model), the ground state lies in the uniform zero-flux sector, while for $p = 0 \mod 4$, it is located in the uniform $\pi$-flux sector.
Lieb's theorem is not restricted to Euclidean lattices. However, it is not applicable to lattices with odd loops, since they do not possess the necessary reflection symmetry. Nevertheless, we conjecture that the ground states lie in the uniform flux sectors, and resort to numerical means in order to validate this conjecture and to characterize the corresponding excitations.

\begin{figure}[b]
    \centering
    \begin{tikzpicture}
        \node[anchor=south west] at (-7, 0){
            \includegraphics[scale=0.8]{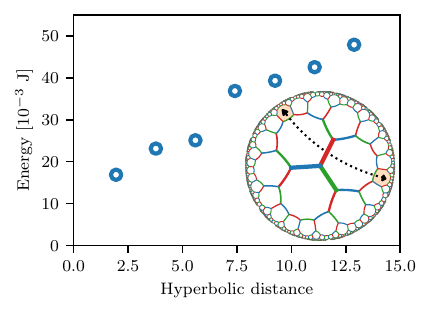}
        };
        \node[anchor=south west] at (0, 0) {
            \includegraphics[scale=0.6]{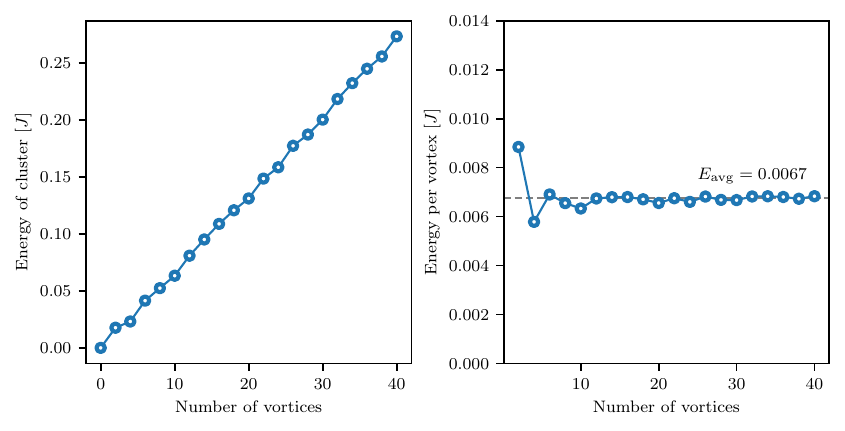}
        };
        \node[anchor=west] at (-7, 4.6) {\textbf{(a)}};
        \node[anchor=west] at (0, 4.6) {\textbf{(b)}};
        \node[anchor=east] at (5, 4.6) {\textbf{(c)}};
    \end{tikzpicture}
    \caption{
        \label{fig:energy_of_vortices}
        \textbf{(a)}
        Energy cost of creating a pair of spatially separated visons in a cluster with 11878 sites for the parameters $J_x = J_y = J_z = \frac{1}{3}$. The energy is increasing with the distance between the visons, hinting at an attractive force between the excitations. 
        \textbf{(b)}
        Energy cost of creating visons and aggregating them in a cluster on a droplet with 7569 sites for the parameters $J_x = J_y = J_z = \frac{1}{3}$. 
        The total energy cost increases approximately linearly with the number of visons.
        \textbf{(c)}
        Energy cost of adding visons to the cluster per vison for the same configuration as in (b).
        The energy cost per vison is positive with respect to the ground state, which means that the configuration with vortices is energetically less favored than the conjectured uniform flux ground state. 
        Note that all results may exhibit finite-size effects, such that the quantitative findings do not necessarily transfer to the thermodynamic limit.
    }
\end{figure}

As a first basic numerical experiment, we explicitly sample every distinct flux configuration of a droplet with 40 sites at the isotropic point $J_x = J_y = J_z$ and find that the energy is minimized only by the uniform $\pm \frac{\pi}{2}$ flux configurations.

Furthermore, we analyze the energy increase due to the creation of vortex excitations (visons) over the assumed ground states in finite-size OBC clusters.
For this, we consider the chiral phase $B$ including the isotropic coupling point, where $J_x = J_y = J_z$. (For an analysis of the dimer phases $A_{\alpha}$, where the dimer limit can be compared to analytical calculations, refer to Section~\ref{appx:toric_code_limit} of the Supplemental Material.)
We compute the energy cost of nucleating a pair of isolated visons in adjacent plaquettes and pulling them apart. 
The result is shown in Figure~\ref{fig:energy_of_vortices}~(a). Although we acknowledge that these results will quantitatively depend on the cluster size (as it is the case in the $A$ phases (c.f.~Sec.~\ref{subappx:dimer_limit_numerical_results})), they indicate qualitatively that there is an attractive force between vortices. This suggests that the formation of vison clusters is energetically favored. 

Thus, the configuration of vison clusters is investigated further: In Fig.~\ref{fig:energy_of_vortices}~(b),~(c) we show the energy cost of adding a single vison to a cluster as well as the total energy of the cluster above the uniform flux sector.
We find that the energy cost per vison is always larger than zero and that the total energy of the cluster increases linearly with the number of visons. 
This result is reasonable, as we expect the energy of the cluster to be proportional to the length of the cluster's boundary, which, in turn, is proportional to the volume of the cluster.
Thus, although these numerical studies are limited to finite-sized samples, all results indicate that the ground state is formed by the Kramers doublet of the flux $\pm \pi/2$ sectors.


\section{Calculation of the Bulk DOS using the Continued Fraction Method}
\label{appx:dos}

\subsection{Continued Fraction Method}

The general idea of the continued fraction (CF) method~\cite{R_Haydock_1972,R_Haydock_1975,PhysRevB.76.180404} for calculating the bulk density of states (DOS) is to compute one element of the resolvent in the local basis, which is equivalent to compute the moments of the local normalized DOS. Let a physical system be described by a tight-binding Hamiltonian $\hat{H}$, and define its resolvent as 
\begin{align}
    \hat{G}(z) = \frac{1}{z - \hat{H}} 
\end{align}
for a complex number $z \in \mathbb{C}$ in the upper half of the complex plane ($\Im[z] > 0$). Furthermore, let $\{\ket{j}\}$ represent the local basis of the single-particle Hilbert space. Then, the local DOS at site $j$ is given by 
\begin{align}
    \rho_{j}(E) = - \frac{1}{\pi} \, \lim_{\eta \rightarrow 0} \, \Im\left[\braket{j | \hat{G}(E + \ii \eta) | j} \right] .
\end{align}
In order to estimate the bulk DOS of the Majorana Hamiltonian~\eqref{eq:SingleParticleHamiltonian} in the hyperbolic plane, we consider the vertex-centered $\{9, 3\}$ lattice, and calculate the local DOS for the central vertex with label $j = 0$. Then, the resolvent can be written in terms of the moments $\mu_{0,r}$ of the Hamiltonian~\cite{R_Haydock_1972} 
\begin{align}\label{eq:resolvent_moments}
    G_{0 0}(z) = \frac{1}{z} \left( 1 + \sum_{r} \frac{\mu_{0, r}}{z^r} \right)\,,
\end{align}
where the $r$-th moment of the local DOS at site $j = 0$ is defined as
\begin{align}\label{eq:moments_hamiltonian}
    \mu_{0, r} = \braket{0 | \hat{H}^r | 0} = \sum_{j_1 , \cdots , j_{r-1}} H_{0 j_1} \, H_{j_1 j_2} \cdots H_{j_{r-1}, 0} \,.
\end{align}
We can interpret this representation of the resolvent as a sum over all loops through the lattice starting and ending at site $0$ with lengths $r = 0, 1, 2, \cdots$. Since the moments of the Hamiltonian $\mu_{0, r}$ at site $0$ are also the moments of the local DOS at site $0$, finding $\mu_{0, r}$ means finding the DOS. In principle, it is necessary to calculate all moments.

In practice, we use the recursion method of Ref.~\cite{R_Haydock_1972} to compute the element of the resolvent. In the following, we restate the procedure: We represent the Hamiltonian in a new basis $\{\ket{\widetilde{n}}\}$, in which it is tridiagonal,
\begin{align}
    \braket{\widetilde{n} | \hat{H} | \widetilde{n}'} 
    = 
    \begin{pmatrix}
        a_0 & \sqrt{b_1} \\ \sqrt{b_1} & a_1 & \sqrt{b_2} \\ & \sqrt{b_2} & a_2 & \ddots \\ & & \ddots & \ddots
    \end{pmatrix} ,
\end{align} 
and for which $\ket{\tilde{0}} = \ket{0}$ holds. Then, the $(0, 0)$-element of the resolvent is given by the \emph{continued fraction}
\begin{subequations}\label{eq:cf_continued_fraction}
    \begin{align}
        G_{0 0}(z) &= \frac{1}{z - a_0 - b_1 \, g_1(z)} \\
        g_1(z) &= \frac{1}{z - a_1 - b_2 \, g_2(z)}    \\
        &\vdots \nonumber \\
        g_n(z) &= \frac{1}{E - a_{n} - b_{n+1} \, g_{n + 1}(z)} .
    \end{align}
\end{subequations}
The coefficients $a_n, b_n \in \mathbb{R}$ in the tridiagonal Hamiltonian representation are calculated using the following recursion: First, we set $\ket{\widetilde{0}} = \ket{0}$, and $b_0 = 0$.
Then we can recursively calculate the following sequence:
\begin{subequations}\label{eq:cf_iteration}
    Given a state $\ket{\widetilde{n}}$ of the new basis, its predecessor $\ket{\widetilde{n - 1}}$ and the square off-diagonal matrix element $b_{n}$ from the previous iteration, we first calculate the diagonal element in the Hamiltonian given by 
    \begin{align}
        a_n = \braket{\widetilde{n} | \hat{H} | \widetilde{n}} \,.
    \end{align}
    Second, we compute the non-normalized auxiliary vector
    \begin{align}
        | \phi_{n + 1} \} = \hat{H} \, \ket{\widetilde{n}}  - b_{n} \, \ket{\widetilde{n-1}} - a_n \, \ket{\widetilde{n}} \,.
    \end{align}
    Then, we find the corresponding coefficient $b_{n + 1}$ as the normalization factor for $| \phi_{n + 1} \}$, \ie 
    \begin{align}
        b_{n + 1} &= || \, |\phi_{n + 1}\} \, ||^2 \,,
    \end{align}
    where $|| \cdot ||$ denotes the Euclidean vector norm, as well as the new normalized state
    \begin{align}
        \ket{\widetilde{n + 1}} &= |\phi_{n + 1}\} / \sqrt{b_{n+1}} \,.
    \end{align}
    With the new state $\ket{\widetilde{n + 1}}$ being determined, the next step in the iteration~\eqref{eq:cf_iteration} can be performed.
\end{subequations}

In principle, this iteration can be repeated indefinitely, and we would obtain all moments $\mu_{0, r}$ of the local DOS. In practice, it is only possible to determine a part of the moments exactly for a finite system. As the length of the loops described by equation~\eqref{eq:moments_hamiltonian} increases, they will reach the edge of the sample and the moments will deviate from the moments of the bulk DOS. This is also the iteration order $N$, after which we terminate the continued fraction~\eqref{eq:cf_continued_fraction}. 

For periodic gapless structures, the coefficients $a_n$ $b_n$ converge slowly towards their respective asymptotic values $a_{\infty}$, $b_{\infty}$, with oscillations associated with van Hove singularities in the spectrum. 
In the ideal case, the coefficients are already sufficiently converged at iteration order $N$, which is the case, for example, for gapless regular hyperbolic tilings~\cite{mosseri2023density}. 

Then, we set $g_{n}(z) \approx g_{\infty}(z)$ for $n > N$, where $g_{\infty}$ is determined by the self-consistent equation
\begin{align}
    g_{\infty}(z) = \frac{1}{z - a_{\infty} - b_{\infty} \, g_{\infty}(z)} 
    \quad 
    \Rightarrow 
    \quad
    g_{\infty}(z) = \frac{z - a_{\infty}}{2 \, b_{\infty}} \left(1 - \sqrt{1 - \frac{4 \, b_{\infty}}{(z - a_{\infty})^2}} \right) .
\end{align}
In the simplest case, where the coefficients are converged, we identify $a_{\infty} = a_{N}$ and $b_{\infty} = b_{N}$. 

For our model, the convergence is much slower than for regular hyperbolic tiling.
Thus, we obtain better results by setting $b_{\infty}$ to the average over a suitable set of coefficients before the cutoff at index $N$ (see Fig.~\ref{fig:phase_diagram_dos}(a)). From the termination term $g_{\infty}(z)$, it becomes clear that the $a_{\infty}$ can be interpreted as the center of the local DOS, while $b_{\infty}$ corresponds to the bandwidth. Note that for our system we find $a_n \equiv 0$, as the spectrum is symmetric with respect to $E = 0$. 

This termination procedure works reasonably well as long as the DOS is not gapped. In the presence of a gap, the termination should be chosen differently~\cite{P_Turchi_1982}. In this case, the coefficients $b_n$ do not converge to a single value, but alternate between two attractors for large $n$ (see Fig.~\ref{fig:phase_diagram_dos}(a) of the main text as well as Ref.~\cite{P_Turchi_1982}). Then, the results become more precise when this behavior is reflected in the termination. For $a_n = 0$, we distinguish $b_{\infty}^{\text{odd}} = \lim_{k \rightarrow \infty} b_{2 k + 1}$ and $b_{\infty}^{\text{even}} = \lim_{k \rightarrow \infty} b_{2 k}$ and set $g_{n}(z) \approx g_{\infty}(z)$ for $n > N$ with $N$ even, such that 
\begin{align}
    g_{\infty}(z) = \frac{1}{z - \frac{b_{\infty}^{\text{odd}}}{z - b_{\infty}^{\text{even}} \, g_{\infty}(z) }}
    \quad\Rightarrow\quad 
    g_{\infty}(z) 
    = 
    \frac{1}{2 \, b_{\infty}^{\mathrm{even}} \, z} 
    \left[ 
        z^2 + b_{\infty}^{\mathrm{even}} - b_{\infty}^{\mathrm{odd}} - \sqrt{\left( z^2 + b_{\infty}^{\mathrm{even}} - b_{\infty}^{\mathrm{odd}} \right)^2 - 4 \, b_{\infty}^{\mathrm{even}} \, z^2}
    \right] \,.
\end{align}
The results for representative points in phase diagram can be found in Figure~\ref{fig:phase_diagram_dos}(a) of the main text. 
Finally, note that the difference $|b_{\infty}^{\mathrm{odd}} - b_{\infty}^{\mathrm{even}}|$ can be used as measure for the band gap. This enables to calculate the phase diagram shown in Fig.~\ref{fig:phase_diagram_dos}(b) of the main text.

\subsection{Comparison with Exact Diagonalization and Hyperbolic Band Theory results}

\begin{figure}
    \centering
    \begin{tikzpicture}
        \node[anchor=south west] at (0, 0) {\textbf{(a)} $J_x = 1/6$, $J_y = J_z = 5/12$};
        \node[anchor=north west] at (0, 0) {
            \includegraphics[scale=0.95]{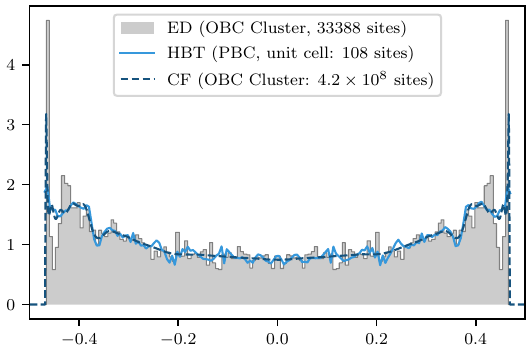}
        };
        \node[anchor=south west] at (9, 0) {\textbf{(b)} $J_x = J_y = J_z = 1/3$};
        \node[anchor=north west] at (9, 0) {
            \includegraphics[scale=0.95]{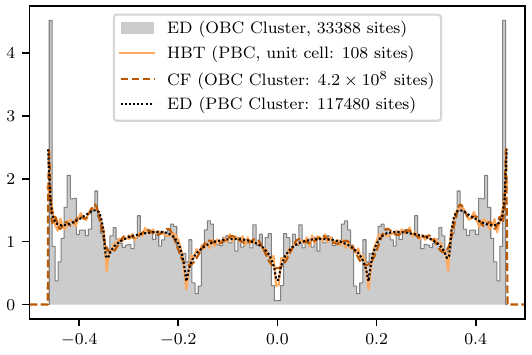}
        };
        \node[anchor=south west] at (0, -6.5) {\textbf{(c)} $J_x = 1/2$, $J_y = J_z = 1/4$};
        \node[anchor=north west] at (0, -6.5) {
            \includegraphics[scale=0.95]{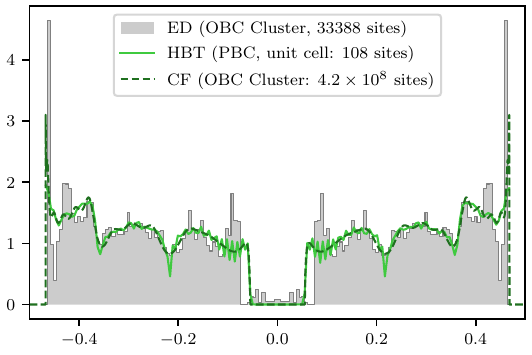}
        };
        \node[anchor=south west] at (9, -6.5) {\textbf{(d)} $J_x = 2/3$, $J_y = J_z = 1/6$};
        \node[anchor=north west] at (9, -6.5) {
            \includegraphics[scale=0.95]{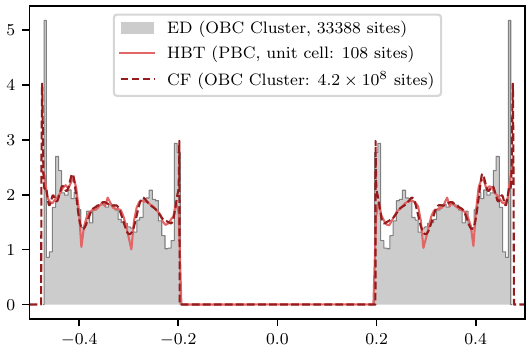}
        };
    \end{tikzpicture}
    \caption{
        \label{fig:dos_comparison}
        Comparison of different methods (Continued fraction method (CF), Exact Diagonalization (ED), and Hyperbolic Band Theory (HBT) with random sampling of momenta) to compute the (bulk) DOS of the Majorana Hamiltonian for the uniform flux sector and different points in the phase diagram. 
        The CF results are computed on a vertex-centered droplet configuration with \num{4.2e6} sites.
        The ED results are computed on a vertex-centered droplet configuration with \num{33388} sites as well as for an PBC cluster with \num{117480} for the isotropic point $J_x = J_y = J_z = \frac{1}{3}$ (panel (b)). 
        For the OBC DOS we include sites within \SI{95}{\percent} of the radius of the Poincaré disk. 
        The HBT results are computed on a compactified graph with \num{108} sites.
    }
\end{figure}

We compare the results obtained by the continued fraction (CF) method with results from Exact Diagonalization (ED) and Hyperbolic Band Theory (HBT) methods. The ED results are obtained by diagonalizing the Majorana Hamiltonian~\eqref{eq:SingleParticleHamiltonian} for a droplet configuration with \num{33388} sites and open boundary conditions (OBC) as well as for periodic boundary conditions (PBC) in the case $J_x = J_y = J_z = \frac{1}{3}$. From the spectrum, we deduce the DOS
\begin{align}
    \rho_{\mathrm{Bulk}} = \sum_{i \in \mathrm{Bulk}} |\psi_{n, i}|^2 \, \delta(E - E_n)
\end{align}
by counting the number of eigenvalues in small energy bins. In order to find the DOS of the bulk for OBC, we weight each eigenstates by its weight in the inner \SI{95}{\percent} percent of the Poincaré disk. 

In order to calculate the spectrum from HBT, we employ the following trick: Since we are only interested in the eigenvalues, we can apply any similarity transform to the Hamiltonian as long as the flux of $\pm \pi/2$ through a plaquette is invariant. With such a mapping, the Majorana Hamiltonian can be transformed into a single-particle Hamiltonian with a constant magnetic field. This problem was already solved for periodic boundary conditions in Reference~\cite{PhysRevLett.128.166402}. For these results we use a compactified hyperbolic graph with \num{108} sites. We calculate its hyperbolic Bloch Hamiltonian and sample the momenta randomly.

Figure~\ref{fig:dos_comparison} compares the resulting DOS of the three methods. We observe that the results agree qualitatively. 
For example, all methods predict the presence of the pronounced local minima in the $J_x = J_y = J_z = 1/3$ case. But there are mismatches in the exact location and the depth of the minima. Most importantly, however, all methods agree that there are no gaps in the spectrum within the chiral phase. 

The deviations arise since all three methods have their drawbacks: The ED methods works only for sufficiently small droplet sizes. Moreover, the categorization of sites belonging to the bulk or to the edge by their position in the Poincaré disk depends on the choice of the cutoff: If it is too small, important contributions to the bulk DOS may be missing, if the cutoff is too large, the DOS picks up edge effects. Furthermore, the results may depend on the binning procedure. 
These issues seem to be mitigated for the PBC cluster. However, such a PBC calculation was only performed for the isotropic point $J_x = J_y = J_z$ where the coloring of the graph does not play a role.
The results using Hyperbolic Band Theory on a compactified graph can also suffer from finite-size effects. Depending on the size of the unit cell, contributions to the DOS coming from higher-order representations of the translational group may be missing. 
The CF method is applicable to much larger droplet sizes, since the method requires only calculating matrix-vector products. Still, only a certain number of moments will be exact, before the continued fraction must be terminated. This can lead to convergence problems, especially close to sharp peaks or dips in the DOS. As we see in Fig.~\ref{fig:dos_comparison}~(b), the CF results exhibit ripple artifacts oscillating around the PBC ED result.

Despite the drawbacks of the individual methods, we conclude that the combination of all three methods is suitable to predict qualitative features of the DOS of the Majorana Hamiltonian on the $\{9, 3\}$ lattice. %


\section{Dimer limit and perturbation theory in the gapped phase}
\label{appx:toric_code_limit}

\subsection{Derivation of the effective Hamiltonian for vison excitations in the dimer limit}
\label{subappx:perturbation_theory}

\begin{figure*}
    \centering
    \begin{tikzpicture}
        \node[anchor=south west] at (0, 0) {
            \includegraphics[scale=.5]{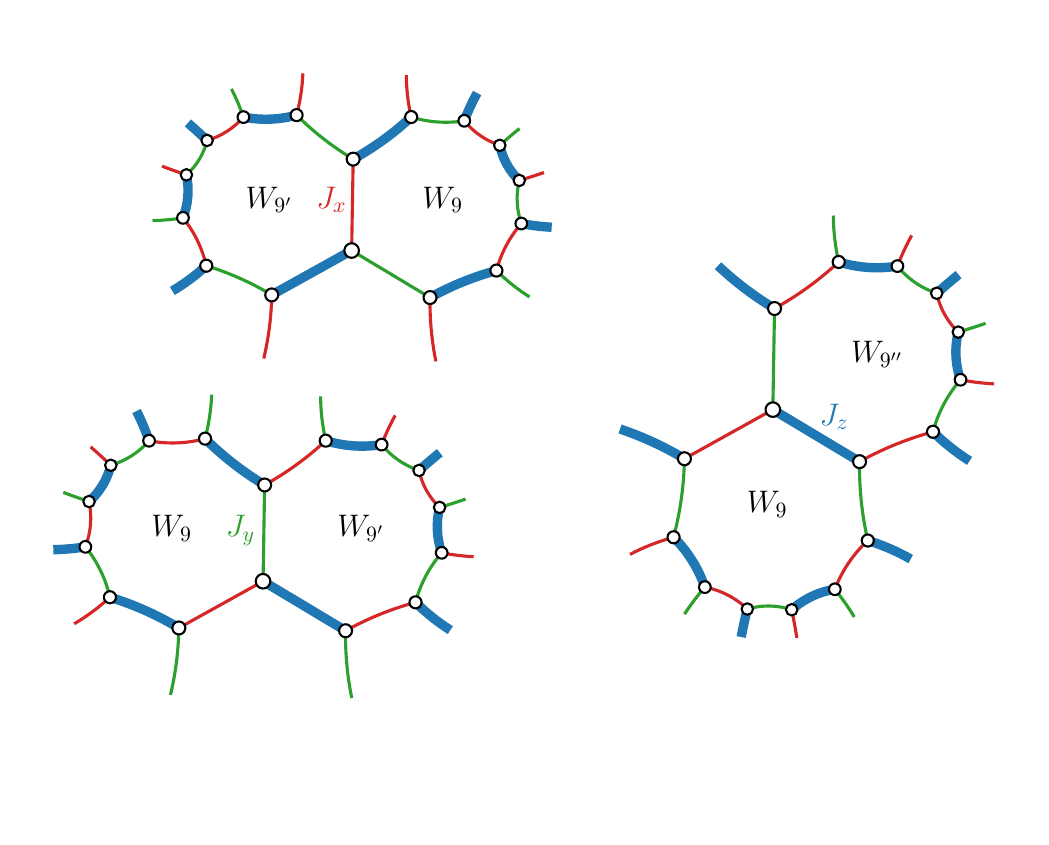}
        };
        \node[anchor=south west] at (9, 1.5) {
            \includegraphics[scale=.4]{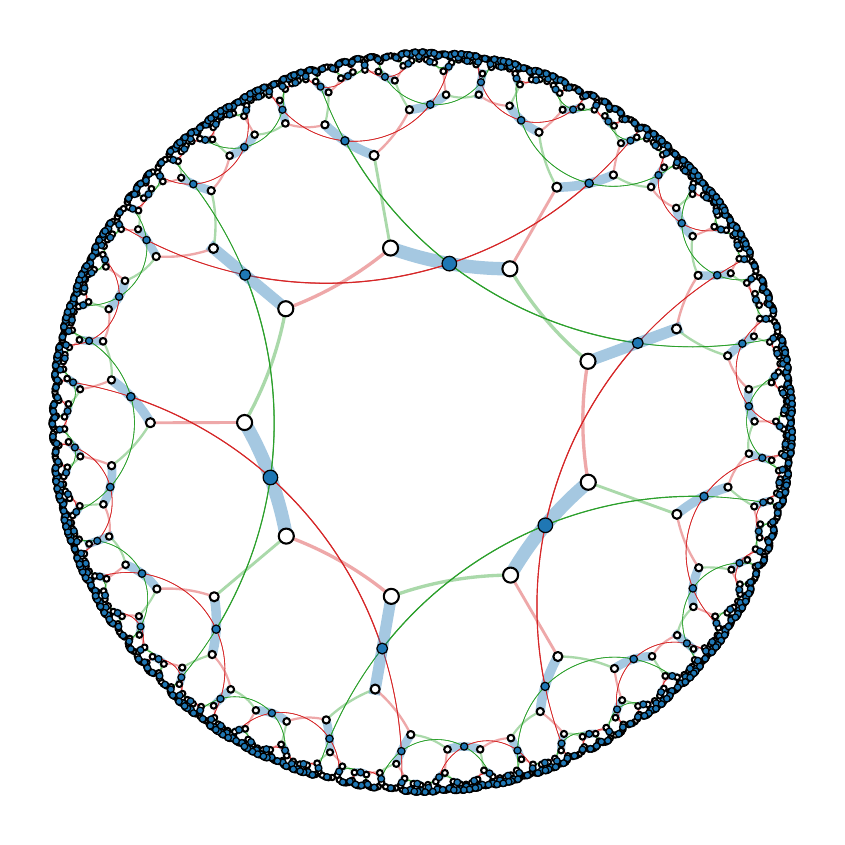}
        };
    
        \node[anchor=west] at (0, 6.5) {\textbf{(a)}};
        \node[anchor=east] at (9, 6.5) {\textbf{(b)}};
    \end{tikzpicture}
        \vspace{-1cm}
\caption{
        \label{fig:tc-limit}
        \textbf{(a)}
        The three lowest order non-self-retracing paths on the $\{9,3\}$ lattice in the limit $J_z \gg J_x, J_y$. The thickness of the links represents the strength of coupling. The two paths on the left contribute to the effective Hamiltonian at 10th order, the path on the right at 12th order.
        \textbf{(b)} In the limit of strong coupling anisotropy $J_z \gg J_x, J_y$, the $\{9,3\}$ Kitaev model lattice maps onto a spin-1/2 model on a $\{6,4\}$ hyperbolic lattice. Figuratively, this corresponds to collapsing the strong links on $\{9,3\}$ into effective sites (blue dots), connected via weak $J_x$ and $J_y$ links (green and blue).
        }
\end{figure*}

The phases $A_x$, $A_y,$ and $A_z$ are realized when one of the couplings constants $J_\alpha$ (with $\alpha = x, y, z$) is larger than the sum of the other two. These phases contain the point where one of the couplings dominates. 
Without loss of generality, we consider $J_z \gg J_x, J_y$ and $J_\alpha > 0\;\forall\alpha$. 
In the limit $J_x = J_y = 0$, the ground state consists of decoupled Ising dimers. If we contract these dimers to a point, the effective lattice becomes a hyperbolic $\{6, 4\}$ lattice (Fig.~\ref{fig:tc-limit}(b)). 
We can investigate the regime around this limit using perturbation theory. 

For this we use the Fermion representation of the problem and the Feynman diagram approach of Petrova, et al.~\cite{Petrova2014}.
Let us re-state the Feynman rules for this analysis in the following: 
\begin{enumerate}[label={(\roman*)}]
    \item 
    Starting from any given vertex, construct all possible paths that terminate at that vertex.
    \item 
    Each link contained in the path gives a contribution (i) $2 J_x u_{jk}$ if $(j,k)$ is a weak link, (ii) $2 J_z u_{jk} / (\omega^2 + 4J_z^2)$ if $(j,k)$ is a strong link and (iii) $\omega/ (\omega^2 + 4J_z^2)$ if $(j,k)$ is a strong link that is attached to the path with only one site. The total amplitude of each path is then obtained by multiplying every contribution and integrating this product over the frequency $\omega$.
    \item 
    The effective Hamiltonian is then given by the sum of the amplitudes of all possible paths, where the reverse of a path is treated as distinct. 
\end{enumerate}

Note that self-retracing paths give flux-independent contributions because each link $(j,k)$ is traversed once in every direction, giving a constant $u_{jk}u_{kj} =-1$. Moreover, the contributions of a path of odd length cancels out with the same path traversed in opposite direction. Thus, only noncontractible even-length paths give nontrivial contributions. 

There are three candidate paths of length 16, two paths along two nonagons sharing a weak $J_x$ or $J_y$ link, and one along the boundary of two nonagons sharing a strong link, respectively, shown in Fig.~\ref{fig:tc-limit}(a). Their contributions are
\begin{subequations}
\begin{align}
    I_{\text{weak},x} 
    &= 
    2 \cdot \int_{-\infty}^{\infty} \frac{d\omega}{2\pi} \, \frac{1}{2} \,
    (2J_x)^4 (2J_y)^6
    \left( \frac{2J_z}{\omega^2+ 4J_z^2}\right)^6 
    \left( \frac{\omega}{\omega^2 + 4J_z^2} \right)^4 W_p W_{p'} 
    = 
    \frac{143 J_x^4 J_y^6}{2^{16} J_z^9} W_p W_{p'} ,
\end{align}
\begin{align}
    I_{\text{weak},y} 
    &= 
    2 \cdot \int_{-\infty}^{\infty} \frac{d\omega}{2\pi} \, \frac{1}{2} \,
    (2J_x)^6 (2J_y)^4
    \left( \frac{2J_z}{\omega^2+ 4J_z^2}\right)^6 
    \left( \frac{\omega}{\omega^2 + 4J_z^2} \right)^4 W_p W_{p'} 
    = 
    \frac{143 J_x^6 J_y^4}{2^{16} J_z^9} W_p W_{p'}
\end{align}
and
\begin{align}
    I_{\text{strong},z} 
    &= 
    2 \cdot \int_{-\infty}^{\infty} \frac{d\omega}{2\pi} \, \frac{1}{2} \,
    (2J_x)^6 (2J_y)^6
    \left( \frac{2J_z}{\omega^2+ 4J_z^2} \right)^4 
    \left( \frac{\omega}{\omega^2 + 4J_z^2} \right)^8 W_p W_{p''} 
    = 
    \frac{91 J_x^6 J_y^6}{2^{19} J_z^{11}} W_p W_{p''} .
\end{align}
\end{subequations}
Note that the factors of $2$ in front of the integrals come from tracing the paths in forward and backward direction.
The $x$ and $y$ contributions are at 10th order, the $z$ contribution is at 12th order. 
Since these contributions only represent the respective leading order, the contributions by the strong links can be discarded to obtain the 10th order effective Hamiltonian.
One finds
\begin{align}\label{eq:effective_ham_appendix}
    \hat{H}_\text{eff}^{(10)} = 
    \text{const}
    + 
    \sum_{\langle p, p' \rangle_{\mathrm{weak},x}} \frac{143 \, J_x^4 \, J_y^6}{2^{16} \, J_z^9} \, \hat{W}_p \hat{W}_{p'} 
    + 
    \sum_{\langle p, p' \rangle_{\mathrm{weak},y}} \frac{143 \, J_x^6 \, J_y^4}{2^{16} \, J_z^9} \, \hat{W}_p \hat{W}_{p'} ,
\end{align}
where we sum over two neighboring plaquettes $p$ and $p'$, adjacent to either a weak $x$-type or $y$-type link.

Regarding the dimer limit, this result confirms our conjecture that the ground state lies within the uniform flux sector: If $W_p = \pm \ii$ for all plaquettes $p$, the contribution of two neighboring plaquettes acquires the sign $(\pm \ii) (\pm \ii) = -1$.  Thus, the perturbative Hamiltonian is minimized for the uniform flux sector.
The resulting model is a hyperbolic version of the Kitaev Toric Code~\cite{breuckmann2017hyperbolic}.

\subsection{Comparison with numerical results and estimatation of finite-size effects}
\label{subappx:dimer_limit_numerical_results}

\begin{figure*}
    \centering
    \begin{tikzpicture}
        \node[anchor=north west] at (0, 0) {
            \includegraphics[width=0.97\linewidth]{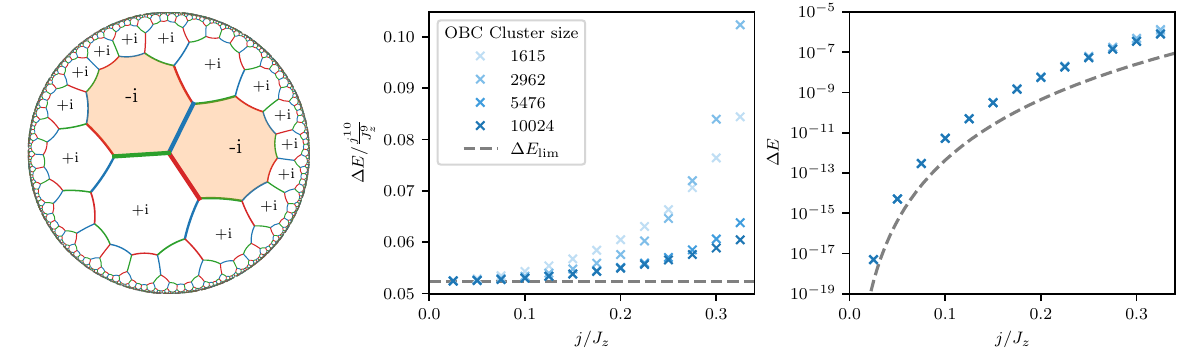}
        };    
        \node[anchor=center] at (0, 0) {\textbf{(a)}};
        \node[anchor=east] at (6, 0) {\textbf{(b)}};
        \node[anchor=east] at (12, 0) {\textbf{(c)}};
    \end{tikzpicture}
        \vspace{-1cm}
\caption{
        \label{fig:dimer_limit_comparison}
        \textbf{(a)}
        OBC cluster with a vison pair created by flipping the $z$ link (blue) between the $W_p = -\mathrm{i}$ plaquettes. The vison pair is surrounded by 6 $x$-type links, 6 $y$-type links, as well as 4 $z$-type links.
        \textbf{(b)/(c)}
        Comparison of the energy of a vison pair computed by diagonalizing the Majorana Hamiltonian for different system size in the limit $J_x = J_y = j \ll J_z$ with the analytical perturbation theory result of the dimer limit.
        Panel (b) shows the data normalized to $(j/J_z)^{10}$ while (c) shows the plain data on a logarithmic scale.
        The numerically obtained energy differences converge towards the analytical result for $j/J_z \rightarrow 0$. Finite-size effects increase with larger $j$.
        }
\end{figure*}

We compare this analytical result to numerical exact diagonalization (ED) calculations for finite-size droplets of different sizes. 
Starting from the uniform $\pi/2$ flux ground state, we create a pair of visons by flipping a strong link in the center of the droplet. This leads to two plaquettes with $W_p = \ii$ surrounded by plaquettes $W_p = -\ii$ (compare Fig.~\ref{fig:dimer_limit_comparison}~(a)). 

Due to the vison pair, the energy of the state is increased with respect to the ground state. If we place ourselves in the dimer limit, i.e. $J_x = J_y = j \ll J_z$, we can compute the energy difference $\Delta E_{\mathrm{lim}}$ to the ground state analytically using equation~\eqref{eq:effective_ham_appendix}. There is an energy contribution for each pair of neighboring plaquettes with opposite flux if the plaquettes are connected by a weak link. Plaquette pairs with the same flux do not contribute, and the corrections from opposite flux plaquettes connected by a strong link are negligible. 
Therefore, we have to consider $12$ contributions in total (6 from $x$ link boundaries to $+\mathrm{i}$ plaquettes, 6 from $y$ link boundaries to $+\mathrm{i}$ plaquettes), leading to an energy increase of
\begin{align}
    \Delta E_{\mathrm{lim}} 
    = 
    2 \cdot 12 \cdot \frac{143 \, J_z}{2^{16}} \left(\frac{j}{J_z}\right)^{10}
    \approx 
    \num{0.052368} \, J_z \, \left(\frac{j}{J_z}\right)^{10}
    \qquad 
    \text{for }
    \frac{j}{J_z} \rightarrow 0
\end{align}
compared to the ground state. 

We check this result numerically by constructing and diagonalizing the Majorana Hamiltonian for this flux configuration. Summing the eigenvalues of the Majorana Hamiltonian in the prescribed way yields the eigenenergy of the many-body state. We compute this energy for the ground state as well as for the two-vison state for different values of $\frac{j}{J_z}$. 
Note that this computation requires caution as the energy difference between both states scales with $(\frac{j}{J_z})^{10}$, which necessitates high computational precision. For example, for $\frac{j}{J_z} = \num{0.1}$, the numerical precision of the eigenvalues must be at least $10^{-13}$ or $10^{-14}$.
The energy difference $\Delta E$ resulting from these computations is compared with the analytical result $\Delta E_{\mathrm{lim}}$ in figure~\ref{fig:dimer_limit_comparison}~(b) and~(c). Panel (b) shows the result normalized to $(\frac{j}{J_z})^{10}$, while panel (c) depicts the actual energy increase on a logarithmic scale. As we can see, the numerical results approach the analytical value as $\frac{j}{J_z} \rightarrow 0$, which indeed confirms that the perturbation theory for the vison excitations in the dimer limit holds.

The numerical data were obtained for different sizes of the OBC cluster (c.f.~Fig.~\ref{fig:dimer_limit_comparison}~(b),~(c)), which enables us to gauge the finite-size effect in this type of calculations. As it turns out, the finite-size effects on excitation energies impact the result significantly for larger values of $\frac{j}{J_z}$. This means especially, that computations at the isotropic point ($j/J_z = 1$) must be assumed to suffer from certain finite-size effects for the available system sizes. This holds certainly also for the quantitative findings of figure~\ref{fig:energy_of_vortices}.


\section{Calculation of the local Chern marker}
\label{appx:chern_marker}

\begin{figure}
    \centering
    \begin{tikzpicture}
        \node[anchor=south west] at (0, 0) {
            \includegraphics[scale=.85]{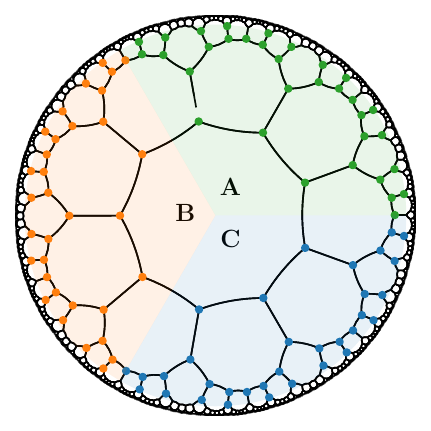}
        };   
        \node[anchor=south west] at (7, -0.4) {
            \includegraphics[scale=.9]{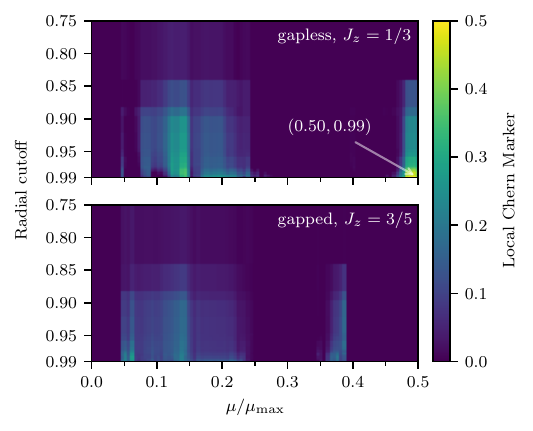}
        };
    \node[anchor=west] at (0, 6) {\textbf{(a)}};
    \node[anchor=east] at (7, 6) {\textbf{(b)}};
    \end{tikzpicture}
    \caption{
        \label{fig:lcms}
        \textbf{(a)} The three regions $A, B$, and $C$ in counterclockwise order we used to compute the local Chern marker. \textbf{(b)} The dependence of the LCM on the radius up to which sites are taken into account and the energy cutoff $\mu$ at the isotropic point (top, $J_x = J_y = J_z = 1/3$) and in the gapped phase (bottom, $J_x = J_y = 1/5$, $J_z = 3/5$). The LCM is symmetric with respect to the axis $\mu/\mu_\text{max}=0.5$. The arrow indicates the parameters we used for computing the LCM phase diagram in Fig.~\ref{fig:phase_diagram_chern}.
        }
\end{figure}

In this section we provide further details on the local Chern marker (LCM)~\cite{KITAEV20062}, which we used in order to study the chiral character of the gapless phase. Let $\ket{\psi_n}$ denote the eigenstate corresponding to energy $E_n$ of the Majorana Hamiltonian in the $-\frac{\pi}{2}$ flux configuration. Furthermore, let $\hat{P}$ be a projector onto all eigenstates with energy below an energy cutoff $\mu$,
\begin{align}
    \hat{P} = \sum_{E_n<\mu} \ket{\psi_n}\bra{\psi_n} \,.
\end{align}
Then, the LCM is defined as
\begin{align}
    C = 12 \pi \i \sum_{j \in \mathrm{A}} \sum_{k \in \mathrm{B}} \sum_{l \in \mathrm{C}} (\hat{P}_{jk}\hat{P}_{kl}\hat{P}_{lj}-\hat{P}_{jl}\hat{P}_{lk}\hat{P}_{kj}),
\end{align}
where $\mathrm{A}$, $\mathrm{B}$, and $\mathrm{C}$ are three bulk regions of sites arranged in counterclockwise order as shown in Fig.~\ref{fig:lcms}(a). The regions do not include the boundary layer. If the regions extended to the boundary, the LCM vanishes. In Fig.~\ref{fig:phase_diagram_chern} of the main text, we calculated the phase diagram LCM for a cutoff energy of $\mu = \mu_\text{max}/2$, where $\mu_{\text{max}} = \max_n E_n$ is the largest eigenenergy. 
For two representative instances of parameters, $J_x = J_y = J_z = 1/3$ as well as $J_z = 3/5$, $J_x = J_y = 1/5$, we investigate how the LCM is influenced by the radial cutoff and energy cutoff. The results are presented in Fig.~\ref{fig:lcms}(b). 

In our calculation, the value of the LCM is not quantized to integer values. 
While the Chern number is expected to be quantized if the Fermi level results in a band gap between two topological bands, the LCM as a computational tool can possibly take other values, especially in the absence of a band gap as in the present case. Furthermore, the LCM calculation can be affected by finite-size and edge effects in the hyperbolic plane.


\section{Time evolution of the Majorana model}
\label{appx:time_evolution}

In order to further investigate the chiral nature of the gapless phase, we consider the time evolution for the Majorana Hamiltonian in the $\pm\frac{\pi}{2}$ flux configuration at the isotropic point $J_x = J_y = J_z = 1/3$. For this, we create a Gaussian wave packet with $\braket{H}=0$ centered at the intersection of the positive $x$ axis with the boundary of the sample,
\begin{align}
    \label{eq:wave-packet}
   \ket{\Psi_\text{init}} =  \mathcal{N} \sum_{\mathrm{Re}(z_j) > 0.98} \exp \left[ \frac{-|1-z_j|^2}{(2\sigma)^2} \right] \ket{j},
\end{align}
where $\mathcal{N}$ is a normalization constant, $\{\ket{j}\}$ represents the local basis of the Majorana Hilbert space, and $\sigma=0.05$ is the width of the packet. $z_j$ denotes the complex plane representation of the coordinate of site $j$. 

\begin{figure*}
    \begin{tikzpicture}
        \node[anchor=north west] at (0, 3.5) {
            \includegraphics[width=0.21\columnwidth]{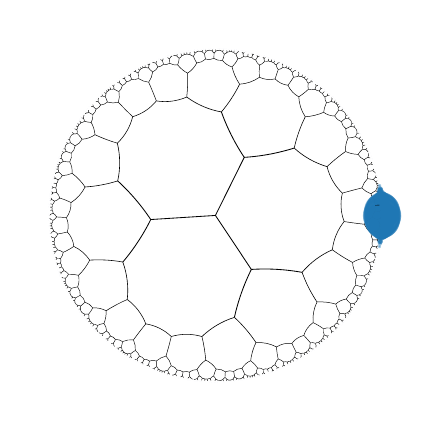}
        };
        \node[anchor=north west] at (3.5, 3.5) {
            \includegraphics[width=0.21\columnwidth]{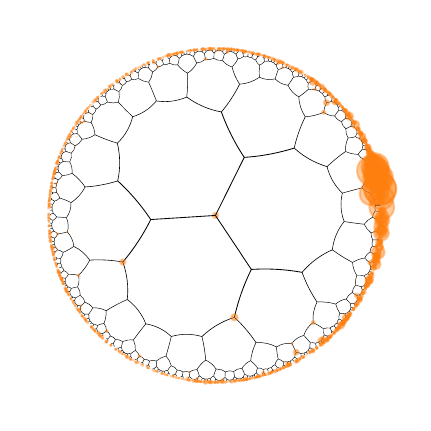}
        };
        \node[anchor=north west] at (7, 3.5) {
            \includegraphics[width=0.21\columnwidth]{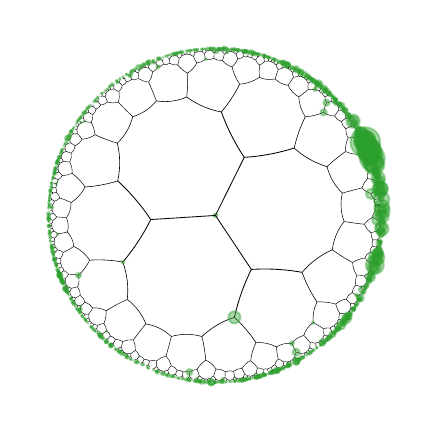}
        };
        \node[anchor=north west] at (10.5, 3.5) {
            \includegraphics[width=0.21\columnwidth]{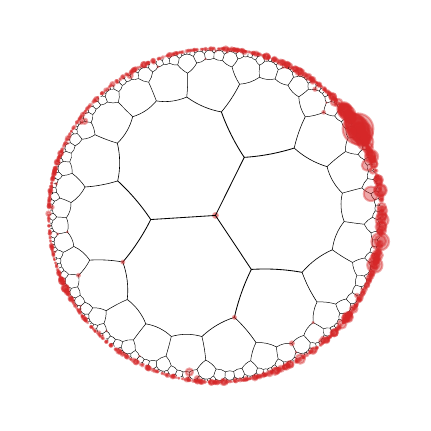}
        };
        \node[anchor=north west] at (14, 3.5) {
            \includegraphics[width=0.21\columnwidth]{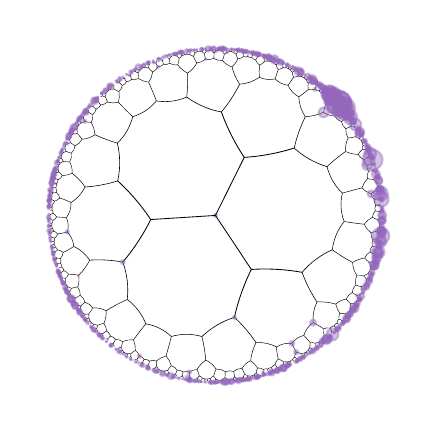}
        };
        
        \node[anchor=north west] at (0, 0) {
            \includegraphics[width=0.21\columnwidth]{Figures/Appendix/edge_propagation-v2_gen_15_+0000.pdf}
        };
        \node[anchor=north west] at (3.5, 0) {
            \includegraphics[width=0.21\columnwidth]{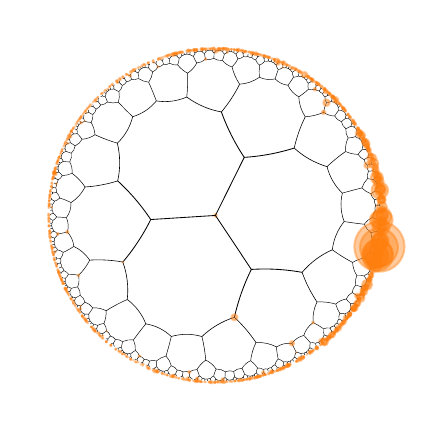}
        };
        \node[anchor=north west] at (7, 0) {
            \includegraphics[width=0.21\columnwidth]{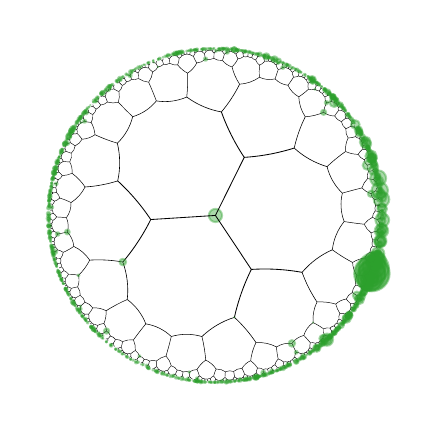}
        };
        \node[anchor=north west] at (10.5, 0) {
            \includegraphics[width=0.21\columnwidth]{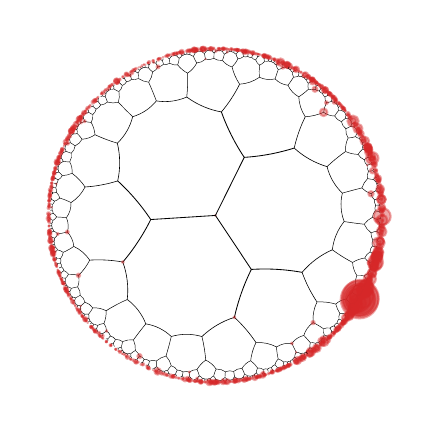}
        };
        \node[anchor=north west] at (14, 0) {
            \includegraphics[width=0.21\columnwidth]{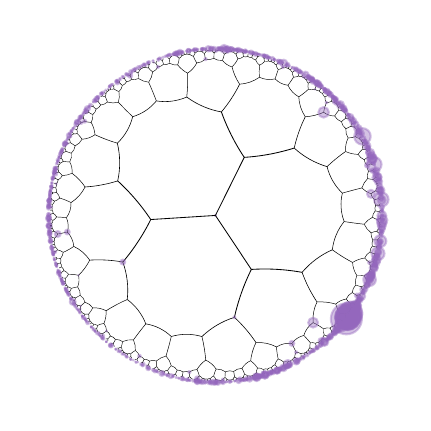}
        };

        \node[anchor=west] at (0.5, 3)     {\textbf{(a)}};
        \node[anchor=west] at (4,   3)     {\textbf{(b)}};
        \node[anchor=west] at (7.5, 3)     {\textbf{(c)}};
        \node[anchor=west] at (11,  3)     {\textbf{(d)}};
        \node[anchor=west] at (14.5,  3)   {\textbf{(e)}};
        
        \node[anchor=west] at (0.5, -.5)   {\textbf{(f)}};
        \node[anchor=west] at (4,   -.5)   {\textbf{(g)}};
        \node[anchor=west] at (7.5, -.5)   {\textbf{(h)}};
        \node[anchor=west] at (11,  -.5)   {\textbf{(i)}};
        \node[anchor=west] at (14.5,  -.5) {\textbf{(j)}};
        
    \end{tikzpicture}
    \caption{
        \label{fig:edge-propagation}
        \textbf{(a)} Initialized wave packet at $t=0$ localized around the rightmost boundary point of the sample in the gapless chiral phase at the isotropic point in the $-\frac{\pi}{2}$ flux sector.
        \textbf{(b)-(e)} Propagation of a wave packet using the time evolution operator $e^{-\ii \hat{H} t}$ and time steps $dt = 800$. The center of the packet clearly propagates counterclockwise along the boundary.
        \textbf{(f)} Same wave packet as in (a) in the $+\frac{\pi}{2}$ flux sector.
        \textbf{(g)-(j)} The packet propagates in the opposite direction with the same angular velocity as in the $-\frac{\pi}{2}$ flux case because both configurations are related via time-reversal symmetry.
    }
\end{figure*}

The time evolution is given by the operator $\exp[-\ii \hat{H} t]$, where $\hat{H}$ is the free Majorana Hamiltonian in the $\pm\frac{\pi}{2}$ flux configuration.
The time evolution of the state $\ket{\Psi_\text{init}}$ on a droplet with 21754 sites is shown in Fig.~\ref{fig:edge-propagation}(a)-(e)\ for the $-\frac{\pi}{2}$ flux sector. We observe that the center of the packet propagates along the boundary counterclockwise. 
Moreover, we see that the wave packet disperses and leaks into the bulk. Since the eigenspectrum of the Hamiltonian is gapless in the chiral phase, the wave packet excitation does not necessarily consist solely of states which are localized at the boundary. Thus, such behavior is expected.
In the $+\frac{\pi}{2}$ flux sector the packet propagates in opposite direction (Fig.~\ref{fig:edge-propagation}(f)-(j)), as these configurations are related via time reversal.    

\end{document}